\documentclass[twocolumn,showpacs,preprintnumbers,amsmath,amssymb,floatfix]{revtex4}
\usepackage{graphicx}% Include figure files
\usepackage{dcolumn}% Align table columns on decimal point
\usepackage{bm}% bold math
\usepackage{color}
\begin{document}

\title{
Dipolar Bosons in Triangular Optical Lattices: \\
Quantum Phase Transitions and Anomalous Hysteresis
}

\author{Daisuke Yamamoto$^{1}$}
\altaffiliation
[Present address:
]{Condensed Matter Theory Laboratory, RIKEN, Wako, Saitama 351-0198, Japan.}
\author{Ippei Danshita$^{2}$}
\author{Carlos A. R. S\'a de Melo$^{3}$}
\affiliation{
{$^1$Department of Physics, Waseda University, Shinjuku-ku, Tokyo 169-8555, Japan}
\\
{$^2$Computational Condensed Matter Physics Laboratory, RIKEN, Wako, Saitama 351-0198, Japan}
\\
{$^3$School of Physics, Georgia Institute of Technology, Atlanta, Georgia 30332, USA}
}
\date{\today}% It is always \today, today,
             % but any date may be explicitly specified
\begin{abstract}
We study phase transitions and hysteresis in a system of dipolar bosons loaded 
into triangular optical lattices at zero temperature. 
We find that 
the quantum melting transition 
from supersolid to superfluid phase 
is first-order, in contrast with the previous report. 
We also find that due to strong quantum fluctuations the supersolid (or solid)-superfluid transition can exhibit an anomalous 
hysteretic behavior, in which the curve of density versus 
chemical potential does not form a standard loop structure. 
Furthermore, we show that the transition occurs 
unidirectionally along the 
anomalous hysteresis curve.
\end{abstract}
\pacs{03.75.-b, 05.30.Jp, 67.80.kb}
\maketitle

%%%%%%%%%%%%%%%%%%%%%%%%%%%%%%%%%%%%%%%%%%%%%%%%%%%%%%%%%%%%%%%%%%%%%%%%%%%%%%%
%%%%%%%%%%%%%%%%%%%%%%%%%%%%%%%%%%%%%%%%%%%%%%%%%%%%%%%%%%%%%%%%%%%%%%%%%%%%%%%
%%                                                                           %%
%% Section I: introduction                                                   %%
%%                                                                           %%
%%%%%%%%%%%%%%%%%%%%%%%%%%%%%%%%%%%%%%%%%%%%%%%%%%%%%%%%%%%%%%%%%%%%%%%%%%%%%%%
%%%%%%%%%%%%%%%%%%%%%%%%%%%%%%%%%%%%%%%%%%%%%%%%%%%%%%%%%%%%%%%%%%%%%%%%%%%%%%%
Ultracold atomic and molecular gases provide very clean and tunable systems 
to study various phenomena in condensed matter physics. 
Due to the remarkable control of physical parameters,
one can simulate the physics of quantum many-body systems 
in regimes inaccessible to solid-state materials. 
Recently, two important developments have taken place in this area. 
First, experimental techniques for the preparation of ultracold gases 
with strong dipole-dipole interactions have been rapidly advancing 
over the last few years. This has been demonstrated by the realization 
of Bose-Einstein condensation (BEC) of $^{52}$Cr atoms which have 
large magnetic dipole moments~\cite{griesmaier-05,lahaye-07} and 
by the creation of heteronuclear (dipolar) polar 
molecules~\cite{ni-08,ospelkaus-09,aikawa-10}.
Secondly, triangular optical lattices of $^{87}$Rb have been created 
experimentally, where the superfluid (SF) to Mott insulator transition
has been observed~\cite{becker-10}.

Stimulated by these experimental developments, we focus on a system of 
ultracold dipolar bosons loaded into a triangular optical lattice. 
Due to its long-range nature, the dipole-dipole interaction coupled 
with the geometry of the triangular lattice 
can produce strong frustration. This setup provides an ideal 
venue for studying the interplay between strong frustration and 
quantum fluctuations. The studies of frustration have been carried out 
mainly in the field of magnetic materials. The frustration of spins 
can lead to exotic low-temperature spin states, such as spin glass, 
spin liquid, and spin ice~\cite{gardner-99,balents-10,bramwell-01}. 

It is well known that a system of lattice bosons with finite-ranged repulsion 
can be mapped, in the hard-core limit, onto a quantum spin-1/2 system with 
an $XXZ$-type anisotropy and a longitudinal magnetic field~\cite{matsuda-70}. 
However, unlike usual 
magnetic materials, the exchange interactions of the mapped spin Hamiltonian are ferromagnetic for the $x$ and $y$ components, which means that the frustration arises only from the coupling between 
the $z$ components.
Therefore the studies on strongly interacting 
bosons on frustrated lattices have great potential to pioneer new 
and intriguing phenomena not found in the regime of real spin systems 
and to provide a deeper understanding of geometrical frustration 
from a new perspective. 
Thus, in this paper, we report the effects 
of quantum fluctuations and geometrical frustration in triangular 
optical lattices of dipolar bosons leading to quantum melting of 
supersolid (SS) or solid phases into SFs and to an anomalous 
unidirectional hysteresis in the density versus chemical potential
phase diagram.

To capture the physics described above, we model dipolar bosons,
for a sufficiently strong on-site interaction, 
by the following hard-core dipolar Bose-Hubbard model on the triangular 
lattice~\cite{pollet-10}: 
\begin{eqnarray}
\hat{H}=
-J\sum_{\langle j,l \rangle}
(\hat{a}^{\dagger}_{j} \hat{a}_{l}+{\rm H.c.})
+\sum_{ j<l }
V_{jl} \hat{n}_{j}\hat{n}_{l}
-\mu \sum_j \hat{n}_j,
\label{hamiltonian}
\end{eqnarray}
where $\hat{a}^{\dagger}_{j}$ is the creation operator of a hard-core boson 
at site $j$, $\hat{n}_j=\hat{a}^{\dagger}_{j} \hat{a}_{j}$, 
$J$ is the hopping amplitude between nearest-neighbor sites, 
and $\mu$ is the chemical potential. We assume that the dipole moments are 
polarized by the external field in 
the direction perpendicular to the lattice plane. In this case, the interaction 
between the dipoles is isotropic and 
can be well approximated by 
$V_{jl}=Vd^3/\left|{\bf r}_{j}-{\bf r}_{l}\right|^3$. 
Here, $d$ is the lattice spacing. 

%%%%%%%%%%%%%%%%%%%%%%%%%%%%%%%%%%%%%%%%%%%%%%%%%%%%%%%%%%%%%%%%%%%%%%%%%%%%%%%
%%%%%%%%%%%%%%%%%%%%%%%%%%%%%%%%%%%%%%%%%%%%%%%%%%%%%%%%%%%%%%%%%%%%%%%%%%%%%%%
%%                                                                           %%
%% Section II: method                                                        %%
%%                                                                           %%
%%%%%%%%%%%%%%%%%%%%%%%%%%%%%%%%%%%%%%%%%%%%%%%%%%%%%%%%%%%%%%%%%%%%%%%%%%%%%%%
%%%%%%%%%%%%%%%%%%%%%%%%%%%%%%%%%%%%%%%%%%%%%%%%%%%%%%%%%%%%%%%%%%%%%%%%%%%%%%%
%
%%
\begin{figure}[t]
\includegraphics[scale=0.23]{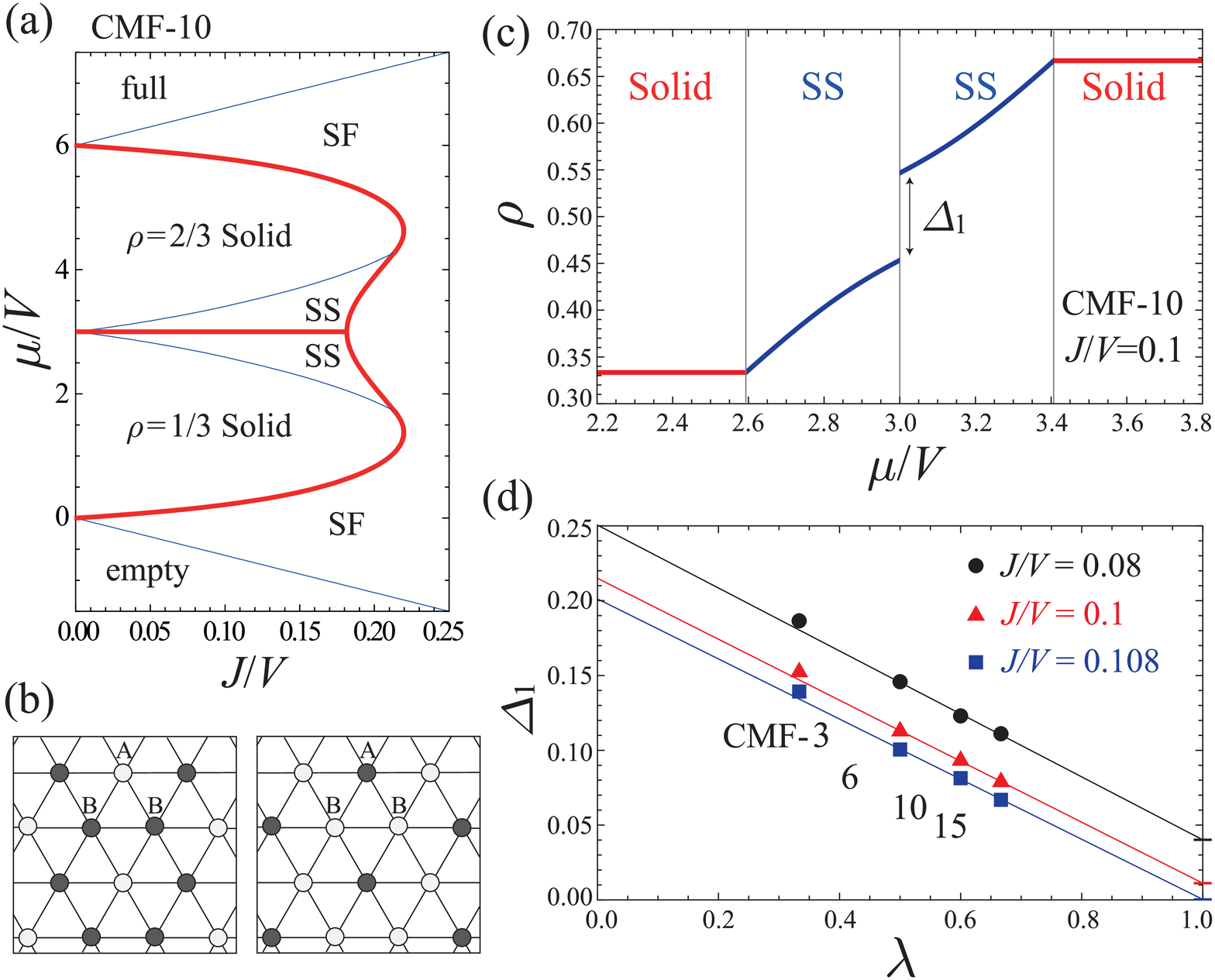}
\caption{\label{fig1} (color online)
(a) The CMF-10 result for the ground-state phase diagram of hard-core bosons with nearest-neighbor repulsion 
on a triangular lattice in the ($J/V,\mu/V$)-plane, where second and first-order phase transitions are indicated by thin blue and thick red lines, respectively. (b) Schematic pictures of the symmetries of 
$\rho=2/3$ solid and upper SS states (left) and of $\rho=1/3$ solid 
and lower SS states (right). The lattice sites are separated into triangular (A) and honeycomb (B) sublattices. 
(c) The filling factor $\rho$ as a function of $\mu/V$ for $J/V=0.1$. 
(d) Cluster-size scaling ($\lambda \rightarrow 1$) of the CMF data for the jump $\Delta_1$ in $\rho$ at $\mu/V=3$. The linear fits of the three points ($N_{\rm C}=6$, $10$, and $15$) are fairly good. 
}
\end{figure}

To study the quantum melting transitions 
and the accompanying 
hysteresis curves, 
we use a cluster mean-field (CMF) method~\cite{oguchi-55,hassan-07,yamamoto-09,EPAPS}, in which we can easily get the 
stationary points of the free energy not only for the globally stable state 
but also for metastable and unstable states. 
 To get meaningful results, one has to employ a sufficiently large cluster for estimating the values of mean fields. 
Although the use of the three-site cluster~\cite{hassan-07} is convenient for tripartite lattices, the size $N_{\rm C}=3$ is too small to see clearly the effects of strong quantum fluctuations. The work described in Ref.~\cite{hassan-07} failed to capture some important features of the phase diagram, e.g., the existence of the direct solid-SF transition at $1/3<\rho<2/3$ and the discontinuity of the SS-SF transition which will be mentioned later. 
Therefore, using much larger-size clusters of triangular shape (up to $N_{\rm C}=15$), here we perform more complex, but more reliable, calculations~\cite{EPAPS}.

%%%%%%%%%%%%%%%%%%%%%%%%%%%%%%%%%%%%%%%%%%%%%%%%%%%%%%%%%%%%%%%%%%%%%%%%%%%%%%%
%%%%%%%%%%%%%%%%%%%%%%%%%%%%%%%%%%%%%%%%%%%%%%%%%%%%%%%%%%%%%%%%%%%%%%%%%%%%%%%
%%                                                                           %%
%% Section III: phase diagram and first-order transitions                    %%
%%                                                                           %%
%%%%%%%%%%%%%%%%%%%%%%%%%%%%%%%%%%%%%%%%%%%%%%%%%%%%%%%%%%%%%%%%%%%%%%%%%%%%%%%
%%%%%%%%%%%%%%%%%%%%%%%%%%%%%%%%%%%%%%%%%%%%%%%%%%%%%%%%%%%%%%%%%%%%%%%%%%%%%%%
To develop some intuition, we consider first hard-core bosons
with only nearest neighbor interactions, i.e., we set $V_{jl} = V$ 
for nearest-neighbor bonds, and $V_{jl}=0$ otherwise. 
The ground-state phase diagram of this simplified model 
contains a wide region of SS phase, 
in which long-range solid (crystalline) order and superfluidity coexist, 
as well as the standard SF and solid phases~\cite{murthy-97,boninsegni-05heidarian-05sen-08heidarian-10,wessel-05}. 
Figure~\ref{fig1}(a) shows the ground-state phase diagram obtained by the ten-site CMF calculation (CMF-10). 
The SF state is characterized by the order parameter 
$\Psi\equiv \sum_j \langle \hat{a}_j\rangle /M$, where $M$ denotes 
the number of lattice sites. The solid states with filling factors
$\rho = 1/3$ and $\rho = 2/3$ have the two-subblatice structures 
depicted in Fig.~\ref{fig1}(b),
which are characterized by 
$\rho_{\bf Q}\equiv \sum_j \langle \hat{n}_j\rangle\exp (i{\bf Q}\cdot {\bf r}_j) /M$ 
with ${\bf Q}=(4\pi/3d,0)$. 
The filling factor is given by $\rho\equiv \sum_j \langle \hat{n}_j\rangle /M$. 
In the SS states, $|\Psi|$ and $|\rho_{\bf Q}|$ have non-zero values simultaneously. 
We determined the boundary lines of first-order transitions from the Maxwell 
construction in ($J/V,\chi$)-plane, 
where 
$
\chi
\equiv 
\sum_{\langle j,l \rangle}
\langle\hat{a}^{\dagger}_{j} \hat{a}_{l}
+
\hat{a}^{\dagger}_{l} \hat{a}_{j}\rangle/M.
$

In Fig.~\ref{fig1}(c) we show the CMF-10 result for the filling factor $\rho$ as a function of $\mu/V$ along the line of $J/V=0.1$.
At the particle-hole symmetry point $\mu/V=3$, we can obviously see a finite jump, $\Delta_1\approx 0.1$ for $J/V=0.1$, in the density curve, whereas previous numerical studies have shown that the density deviation $|\rho-1/2|$ at $\mu/V=3$ is extremely small or even undetectable~\cite{boninsegni-05heidarian-05sen-08heidarian-10,wessel-05}. 
To examine this quantitative difference, we perform the infinite-size extrapolation of the results for different-size clusters [see Fig.~\ref{fig1}(d)] with the scaling parameter $\lambda$ defined by $N_{\rm B}/(N_{\rm C}\times z/2)$, where $N_{\rm B}$ is the number of bonds within the cluster and $z=6$ (see Ref.~\cite{EPAPS}). The jump $\Delta_1$ decreases with cluster size as expected. For example for $J/V=0.1$, its value is strongly reduced from $\Delta_1\approx 0.1$ of CMF-10 to $\Delta_1\approx 0.01$ in the limit $\lambda\rightarrow 1$. The extrapolated value vanishes at $J/V=0.108$, which means that the location of the triple point of the two SS and SF phases is shifted to $(J/V,\mu/V)=(0.108,3)$ in $\lambda\rightarrow 1$.

\begin{figure}[tb]
\includegraphics[scale=0.28]{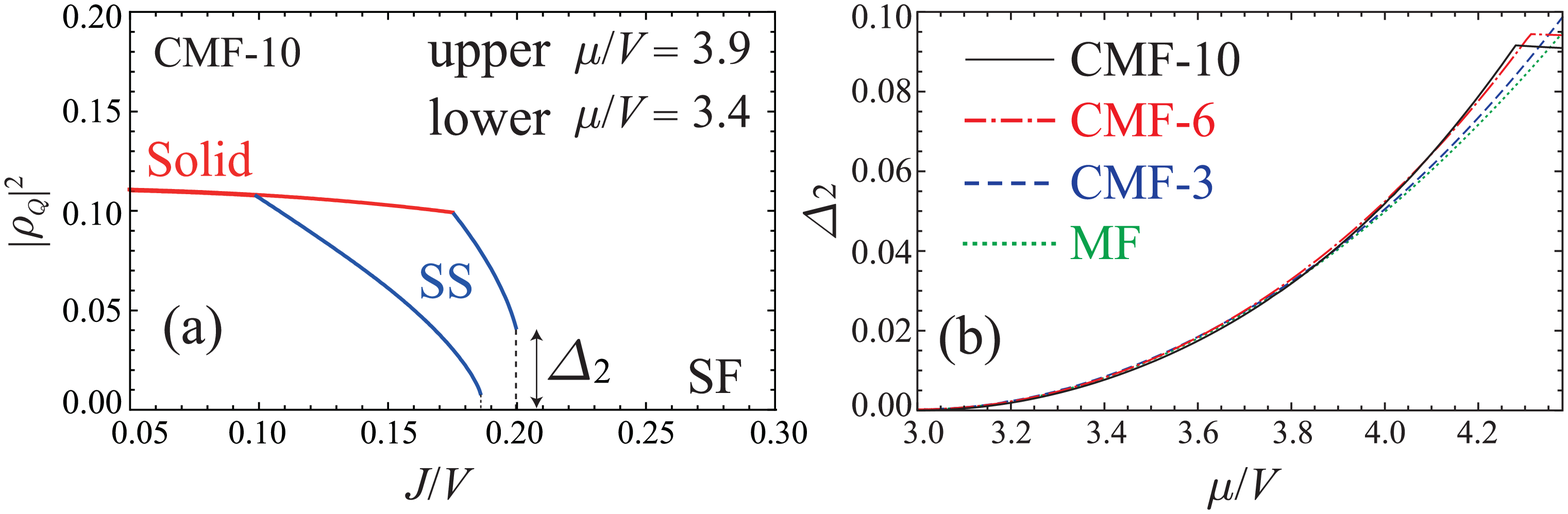}
\caption{\label{fig2} (color online) 
(a) The $J/V$ dependence 
of $|\rho_{\bf Q}|^2$ at $\mu/V=3.4$ and $\mu/V=3.9$. 
(b) The jump $\Delta_2$ in $|\rho_{\bf Q}|^2$ along the SS-SF (or solid-SF) transition line as a function of $\mu/V$. 
}
\end{figure}
The features of the phase diagram in Fig.~\ref{fig1}(a)
are in excellent agreement with those of QMC calculations~\cite{wessel-05}. 
However, there is one major qualitative difference; our CMF results show that 
the transition between SF and SS is first-order (discontinuous). 
To confirm this, we plot $|\rho_{\bf Q}|^2$ as a function of $J/V$, calculated by CMF-10 along the horizontal lines $\mu/V =3.4$ and $\mu/V =3.9$, in Fig.~\ref{fig2}(a). 
For the both values of $\mu/V$, there is a finite discontinuous jump $\Delta_2$ at the SS-SF transition, although that of the former is very small. 
The magnitude of jump decreases monotonically as the value of 
$\mu/V$ approaches the particle-hole symmetry point. It vanishes at $\mu/V=3$. As shown in Fig.~\ref{fig2}(b), the curve shows very little change with increasing cluster size, and thus we conclude that the SS-SF transition is {\it first-order} except for the critical point $\mu/V = 3$. 
In contrast, 
this transition seems to be continuous in QMC simulations, 
see Fig.~8 of Ref.~\cite{wessel-05}, where the authors concluded that 
the SS-SF transition is {\it second-order}. 
This discrepancy may come from a finite-size effect of
QMC calculations, since the magnitude of the jump $\Delta_2$ near $\mu/V = 3$ 
is quite small as shown in Fig.~\ref{fig2}(a). 
For values of $\mu/V$ farther away from the particle-hole symmetry line $\mu/V = 3$, the discontinuous behavior is more pronounced, as shown in Fig.~\ref{fig2}(b), and can now be observed also within QMC~\cite{Note}.

%%%%%%%%%%%%%%%%%%%%%%%%%%%%%%%%%%%%%%%%%%%%%%%%%%%%%%%%%%%%%%%%%%%%%%%%%%%%%%%
%%%%%%%%%%%%%%%%%%%%%%%%%%%%%%%%%%%%%%%%%%%%%%%%%%%%%%%%%%%%%%%%%%%%%%%%%%%%%%%
%%                                                                           %%
%% Section IV: hysteresis                                                    %%
%%                                                                           %%
%%%%%%%%%%%%%%%%%%%%%%%%%%%%%%%%%%%%%%%%%%%%%%%%%%%%%%%%%%%%%%%%%%%%%%%%%%%%%%%
%%%%%%%%%%%%%%%%%%%%%%%%%%%%%%%%%%%%%%%%%%%%%%%%%%%%%%%%%%%%%%%%%%%%%%%%%%%%%%%

\begin{figure}[t]
\includegraphics[scale=0.275]{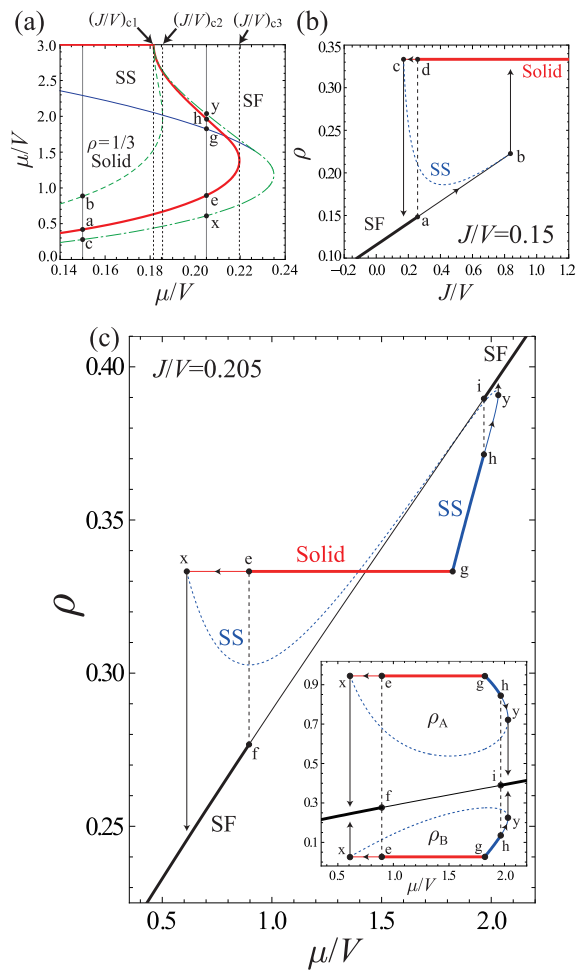}
\caption{\label{fig3} (color online) (a) Magnified view of the 
low-density regions of Fig.~\ref{fig1}(a). 
We plot the limits of metastability of the SF phase 
(dashed green line) and the SS or solid phase 
(dash-dotted green line) in addition to the phase transition lines 
(thin blue and thick red lines) already shown in Fig.~\ref{fig1}(a). 
(b) The usual hysteresis-loop and (c) anomalous unidirectional hysteresis behaviors. The thick solid, thin solid, and dashed lines represent ground, metastable, and unstable states, respectively. 
The inset in panel (c) shows the average density on each sublattice 
denoted by $\rho_{\rm A}$ and $\rho_{\rm B}$, which satisfy 
the relation $\rho=(\rho_{\rm A}+2\rho_{\rm B})/3$. 
The corresponding points in the figures are marked with 
the same letters. }
\end{figure}

Let us discuss the hysteresis in the cycle of decreasing and increasing 
the chemical potential $\mu/V$. The system exhibits different 
hysteretic behaviors in three different ranges of $J/V$, 
which are defined by the thresholds 
$(J/V)_{\rm c1}\approx 0.182$, $(J/V)_{\rm c2}\approx 0.186$, 
and $(J/V)_{\rm c3}\approx 0.220$ 
[marked by the dashed vertical lines in Fig.~\ref{fig3}(a)]. 
In the first region, $J/V<(J/V)_{\rm c1}$, accompanying the 
SF-solid transition, a typical hysteresis loop is formed 
in the ($\mu/V,\rho$)-plane as indicated by the arrows 
in Fig~\ref{fig3}(b). This is simply analogous to a conventional 
liquid-solid transition. In the second region, 
$(J/V)_{\rm c1}<J/V<(J/V)_{\rm c2}$, another hysteresis loop 
is formed around the SS-SF first-order transition point in addition 
to the loop around the SF-solid transition.

Of particular interest is the third region, 
$(J/V)_{\rm c2}<J/V<(J/V)_{\rm c3}$, 
in which the hysteresis exhibits an $anomalous$ behavior. As an example, 
we show in Fig.~\ref{fig3}(c) the solution curves of the CMF-10 
self-consistent equation in the ($\mu/V,\rho$)-plane for $J/V=0.205$. 
There are two first-order transitions, namely, between the solid (at point e) 
and SF (f) states and between the SS (h) and SF (i) states. 
Although the solution branches corresponding to metastable SF and unstable SS 
states apparently cross, the two states at the intersection are {\it not} 
identical. This is clearly seen in the inset of Fig.~\ref{fig3}(c), 
where we plot the sublattice densities. Thus the solution curves are 
completely separated into the line of SF solutions and the twisted 
closed curve consisting of solid and SS solutions. 
In this case,  we have an irreversible quantum melting
transition, and once the solid order is melted
at $T = 0$, it will remain melted. In this regime, the solid phases can 
be reached again only through thermal cycling.

To illustrate this, let us assume that the system is initially 
in a stable solid state located between points e and g in Fig.~\ref{fig3}(c). 
When decreasing $\mu/V$, although a SF state becomes energetically favorable below point e, the solid state remains metastable until it reaches point x. Below point x, the system is destabilized into the true ground state (namely the SF state). 
If we increase $\mu/V$ from the solid state, the system first undergoes a continuous transition 
to the SS phase at point g, and then the metastable SS state is also destabilized into the SF state at point y. On the other hand, the situation drastically changes 
if we start from an initial state in the SF phase. 
We see in Fig.~\ref{fig3}(c) that the globally stable SF solutions 
at low and high $\mu/V$ are connected by the line of metastable 
SF solutions, which means that the SF state is stable for $any$ $\mu/V$. 
Therefore, when we decrease or increase $\mu/V$ starting from a SF state, 
the system remains in the SF phase even if the value of $\mu/V$ enters 
the region where a SS or solid state has the lowest energy. 
Thus the transition in 
varying $\mu/V$ occurs only from the SS (or solid) to SF phase, 
and the hysteresis curve does not form a standard loop structure. 
Crucially, this behavior gets more pronounced and the region $(J/V)_{\rm c2}<J/V<(J/V)_{\rm c3}$ becomes wider as the cluster size $N_{\rm C}$ increases, which means the robustness of the anomalous hysteresis when $\lambda \rightarrow 1$. 

%%%%%%%%%%%%%%%%%%%%%%%%%%%%%%%%%%%%%%%%%%%%%%%%%%%%%%%%%%%%%%%%%%%%%%%%%%%%%%%
%%%%%%%%%%%%%%%%%%%%%%%%%%%%%%%%%%%%%%%%%%%%%%%%%%%%%%%%%%%%%%%%%%%%%%%%%%%%%%%
%%                                                                           %%
%% Section V: full long-range interaction                                    %%
%%                                                                           %%
%%%%%%%%%%%%%%%%%%%%%%%%%%%%%%%%%%%%%%%%%%%%%%%%%%%%%%%%%%%%%%%%%%%%%%%%%%%%%%%
%%%%%%%%%%%%%%%%%%%%%%%%%%%%%%%%%%%%%%%%%%%%%%%%%%%%%%%%%%%%%%%%%%%%%%%%%%%%%%%

%
%%
\begin{figure}[t]
\includegraphics[scale=0.31]{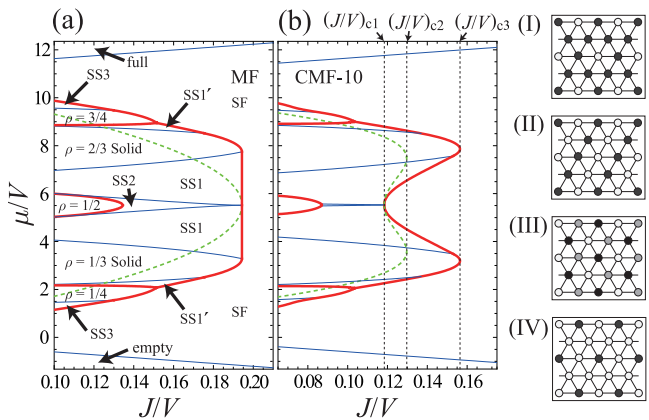}
\caption{\label{fig4} (color online) Ground-state phase diagram of hard-core 
bosons with full long-range dipole-dipole interactions on a 
triangular lattice in the ($J/V,\mu/V$)-plane, obtained from
(a) MF theory and (b) CMF-10 method. Second and first-order 
phase transitions are indicated by thin blue and thick red lines, 
respectively. The dashed green lines represent the limit of metastability 
of the SF phase. The SS1 and SS1' phases have the same sublattice structure 
as the SS phase in Fig.~\ref{fig1}(b). 
(I-IV) Sublattice structures of the $\rho=3/4$ solid and 
the nearby SS3 (I), of the $\rho=1/2$ solid (II), of the SS2 (III), 
and of the $\rho=1/4$ solid and the nearby SS3 (IV). }
\end{figure}
We also discuss the influence of the full long-range dipole-dipole 
interactions~\cite{sansone-10danshita-10} on our results. 
First, we perform a simple mean-field (MF) analysis 
for Eq.~(\ref{hamiltonian}) with 
$V_{jl}=Vd^3/\left|{\bf r}_{j}-{\bf r}_{l}\right|^3$. 
As in the case of the nearest-neighbor interaction model, 
the MF phase diagram depicted in Fig.~\ref{fig4}(a) includes 
large regions of $\rho=1/3$ and $\rho=2/3$ solids and 
the two-sublattice SS phase (named SS1) located between them. 
This is consistent with recent QMC calculations~\cite{pollet-10}. 
Furthermore, the two-sublattice SS phase is stabilized 
also in the regions $\rho<1/3$ and $\rho>2/3$ (SS1'), 
and we find additional SS phases (SS2 and SS3) and solid phases 
with $\rho=1/2$, $\rho=1/4$, and $\rho=3/4$ within the parameter range 
of Fig.~\ref{fig4}(a). To represent these phases, we allowed for 
three- and four-sublattice structures in minimizing the MF energy. 
Next, using the CMF-10 method we take into account quantum fluctuations around the MF solutions [see Fig~\ref{fig4}(b)]. As in the case of nearest-neighbor interactions [Fig~\ref{fig1}(b)], a noteworthy 
consequence of quantum fluctuations is that the SS1-SF boundary 
forms a dip around the particle-hole symmetry line. Moreover, 
a similar anomalous hysteresis described in the nearest-neighbor case
also emerges here, and it is associated with the presence of the dip when 
$(J/V)_{\rm c2}<J/V<(J/V)_{\rm c3}$ 
[$(J/V)_{\rm c2}\approx 0.130$ and $(J/V)_{\rm c3}\approx 0.156$].

Actual experiments of ultracold gases are performed in the presence of 
a trap potential, e.g., $V_{t}({\bf r})=m\omega^2|{\bf r}|^2/2$. 
Within the local-density approximation (LDA), the effective local chemical 
potential is given by $\tilde{\mu}_j=\mu-V_{t}({\bf r}_j)$. 
We suggest that the anomalous hysteretic behavior can be confirmed 
by controlling (decreasing and increasing) $\tilde{\mu}_j$ at the trap 
center via manipulation of, e.g., the frequency $\omega$ 
of the harmonic trap confining the dipolar gases. However, further analyses beyond LDA are still required to fully understand the behavior of the anomalous hysteresis in a trapped system.

%%%%%%%%%%%%%%%%%%%%%%%%%%%%%%%%%%%%%%%%%%%%%%%%%%%%%%%%%%%%%%%%%%%%%%%%%%%%%%%
%%%%%%%%%%%%%%%%%%%%%%%%%%%%%%%%%%%%%%%%%%%%%%%%%%%%%%%%%%%%%%%%%%%%%%%%%%%%%%%
%%                                                                           %%
%% Section VI: summary                                                       %%
%%                                                                           %%
%%%%%%%%%%%%%%%%%%%%%%%%%%%%%%%%%%%%%%%%%%%%%%%%%%%%%%%%%%%%%%%%%%%%%%%%%%%%%%%
%%%%%%%%%%%%%%%%%%%%%%%%%%%%%%%%%%%%%%%%%%%%%%%%%%%%%%%%%%%%%%%%%%%%%%%%%%%%%%%

In summary, we have studied phase transition phenomena in a system of 
dipolar Bose gases loaded into a triangular optical lattice. Using a CMF 
method, we have found that the first-order transition between the 
SS (or solid) and SF phases can exhibit an anomalous hysteretic behavior: 
in varying the chemical potential, the standard hysteresis loop structure 
does not appear, and the phase transition occurs only from the SS (or solid) 
to SF state. This unidirectional 
character is not predicted within the MF (classical) approach, 
since the boundary of the SS-SF transition is given by a 
straight line~\cite{murthy-97}. Moreover, previous studies on 
a similar 
hard-core boson model with nearest-neighbor interactions 
for a square lattice 
have given only a standard hysteresis-loop behavior~\cite{batrouni-00}. 
Thus, the anomalous feature of the hysteresis 
in this system 
is attributed to the interplay between quantum fluctuations and 
the competition of interactions due to the frustrated geometry of 
the triangular lattice.

More specifically, the most important point for the anomalous hysteresis behavior to emerge is the existence of the first-order SF-solid-(SS-)SF transition under varying $\mu/V$, as can be seen from Figs.~\ref{fig3}(a) and (c). Thus we expect to find analogous anomalous hysteresis in a wide range of systems exhibiting $re$-$entrant$ first-order transitions (unless the system has no special symmetry point, e.g., the Heisenberg point in the square-lattice case~\cite{batrouni-00}). 
Recently, it has been suggested that $a$ $spin$ $supersolid$ $state$ can be realized in some magnetic systems~\cite{mag}. Thus, by analogy, we also expect that such quantum spin systems will exhibit a similar anomalous hysteresis behavior as a function of magnetic field.

%%%%%%%%%%%%%%%%%%%%%%%%%%%%%%%%%%%%%%%%%%%%%%%%%%%%%%%%%%%%%%%%%%%%%%%%%%%%%%%
%%%%%%%%%%%%%%%%%%%%%%%%%%%%%%%%%%%%%%%%%%%%%%%%%%%%%%%%%%%%%%%%%%%%%%%%%%%%%%%
%%                                                                           %%
%% Section VI: acknowledgements                                              %%
%%                                                                           %%
%%%%%%%%%%%%%%%%%%%%%%%%%%%%%%%%%%%%%%%%%%%%%%%%%%%%%%%%%%%%%%%%%%%%%%%%%%%%%%%
%%%%%%%%%%%%%%%%%%%%%%%%%%%%%%%%%%%%%%%%%%%%%%%%%%%%%%%%%%%%%%%%%%%%%%%%%%%%%%%
We thank Grant-in-Aid from JSPS (D.Y.), 
KAKENHI (22840051) from JSPS (I.D.)
and ARO (W911NF-09-1-0220) (C.S.d.M.) for support.

\subsection{\large Supplementary Material for ``Dipolar Bosons in Triangular Optical Lattices: \\
Quantum Phase Transitions and Anomalous Hysteresis''}
\renewcommand{\thesection}{\Alph{section}}
\renewcommand{\thefigure}{S\arabic{figure}}
\renewcommand{\thetable}{S\Roman{table}}
\setcounter{figure}{0}
\subsection{\label{1}A. A large-size CMF method and cluster-size scaling} 
The CMF method is convenient and effective in understanding the physics of ordered states including metastability phenomena since all stationary points of the free energy can be obtained not only for the globally stable solution. Moreover, this method is free from ``finite-size effects'' and ``error bars.'' However, in order to get reasonable results, one has to treat sufficiently large clusters as a reference system, especially when strong fluctuations exist in the system. 
In this work, we have used a series of clusters of $N_{\rm C}=3$, 6, 10, and 15 sites given in Table~\ref{table1}. 
%%%%%%%%%%%%%%%%%%%%%%%%%%%%%%%%%%%%%%%%%%%%%%%%%%%%%%%%%%%%%%%%%
\begin{table}[b]
\caption{\label{table1}A series of clusters used in our calculations. The values of $N_{\rm C}$, $N_{\rm B}$, and $\lambda$ are also listed. }
%\begin{ruledtabular}
\begin{tabular}{cccc}
\hline
\hline
~~~~ & MF&CMF-3 & CMF-6 ~~~~ \\ 
\hline \\[-7pt]
~~~~ & \includegraphics[scale=0.3]{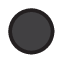}
&\includegraphics[scale=0.3]{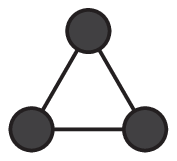}
&\includegraphics[scale=0.3]{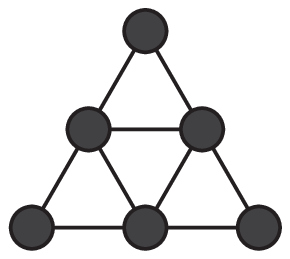}

 ~~~~ \\ \hline 
~~~~$N_{\rm C}$ & 1& 3 & 6 ~~~~  \\
~~~~$N_{\rm B}$ & 0& 3 & 9 ~~~~  \\
~~~~$\lambda$  & 0& 1/3 & 1/2  ~~~~\\
\hline \\[-7pt]
\hline
~~~~ &  CMF-10 & CMF-15 &~~ Exact~~~~ \\ 
\hline \\[-7pt]
~~~~ &\includegraphics[scale=0.3]{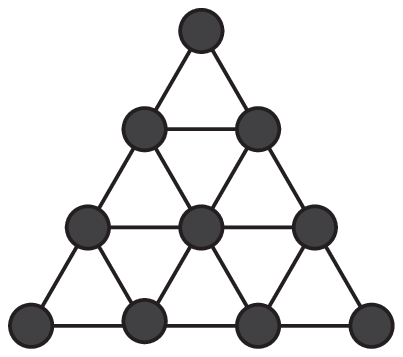}
&\includegraphics[scale=0.3]{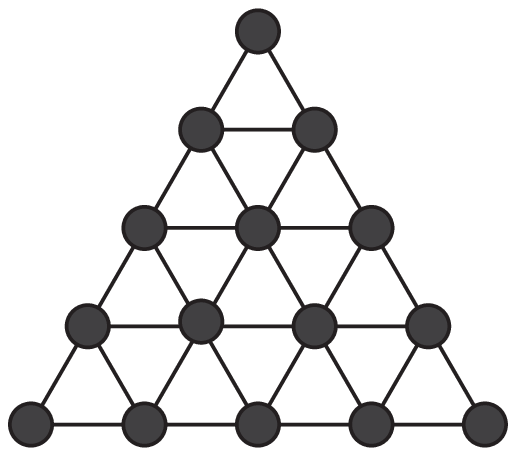}
&~~~~\includegraphics[scale=0.25]{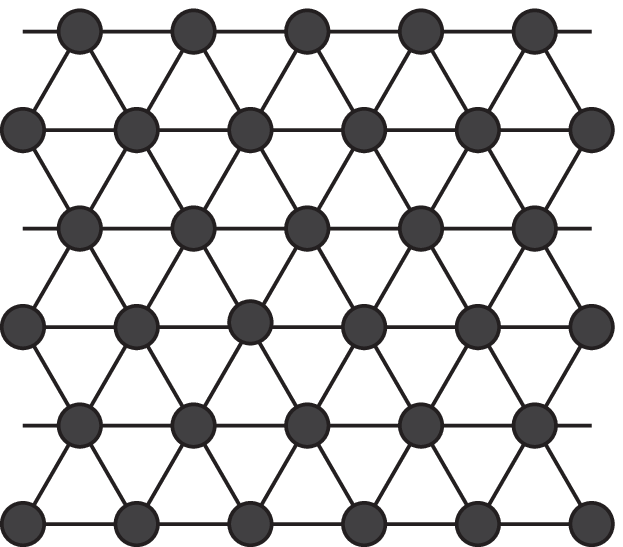}
~~~~ \\ \hline 
~~~~$N_{\rm C}$ &  10 & 15 & ~~~$\infty$~~~~  \\
~~~~$N_{\rm B}$ &  18 & 30 & ~~~$\infty$~~~~  \\
~~~~$\lambda$  &  3/5 & 2/3 &~~ 1 ~~~~\\
\hline
\hline 

\end{tabular}
%\end{ruledtabular}
\end{table}
%%%%%%%%%%%%%%%%%%%%%%%%%%%%%%%%%%%%%%%%%%%%%%%%%%%%%%%%%%%%%%%%%

For a large cluster, we cannot calculate the value of free energy of the system directly from the CMF formalism, unlike the standard MF theory. 
This is a important problem for the system considered in this work, which exhibits first-order phase transitions in a wide region of parameters. 
To overcome this problem, we have used the Maxwell construction in ($J/V,\chi$)-plane to determine the phase boundary. The quantity $\chi$ is defined by
$
\chi
\equiv 
\sum_{\langle j,l \rangle}
\langle\hat{a}^{\dagger}_{j} \hat{a}_{l}
+
\hat{a}^{\dagger}_{l} \hat{a}_{j}\rangle/M,
$ 
and the energy difference between the states at $J_0$ and $J_1$ (when $\mu$ and $V$ are fixed) is given by 
$
-\int_{J_0}^{J_1}\chi (J)d J.
\label{EnergyDifference}
$
Note that the Maxwell construction in ($\mu/V,\rho$)-plane~\cite{batrouni-00a} is inapplicable in this case because of, indeed, the existence of the $anomalous$ hysteresis; as shown in Fig.~3(c), the solution curves are separated into two groups for $(J/V)_{\rm c2}<J/V<(J/V)_{\rm c3}$, and thus one cannot estimate the energy difference between states of different groups by using the integration of the density $\rho$ over the chemical potential $\mu/V$.

\begin{figure}[b]
\includegraphics[scale=0.4]{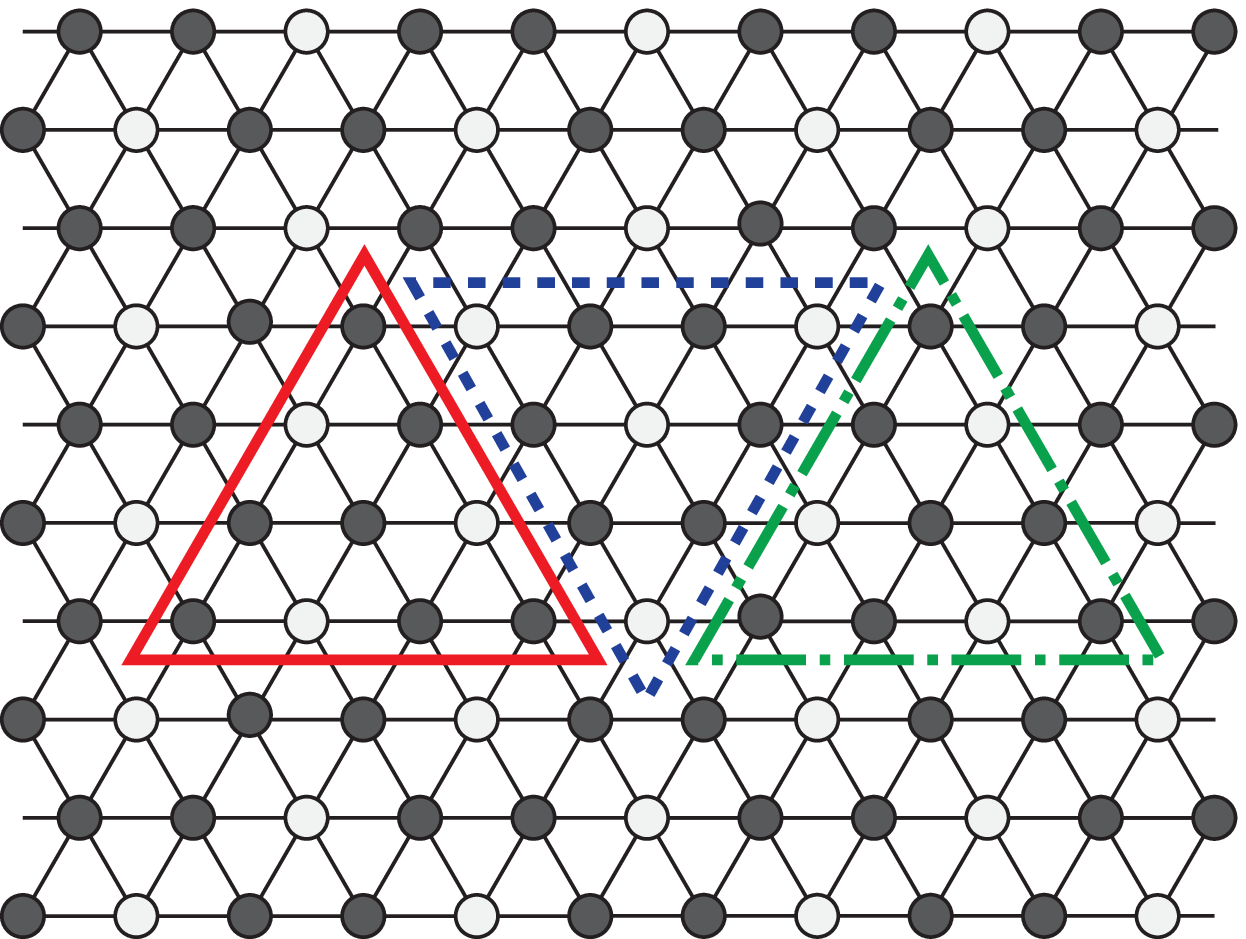}
\caption{\label{figS1}
Schematic picture of the ten-site clusters embedded in the background two-sublattice $\sqrt{3}\times\sqrt{3}$ structure. 
}
\end{figure}

Moreover, to estimate the expectation values
$\langle \cdots \rangle$, one should take the average not only over all internal sites within the cluster, but also all possible choices of how to embed the cluster itself in the background sublattice structure. 
For example, when we assume the two-sublattice $\sqrt{3}\times\sqrt{3}$ 
ordering in the ten-site CMF approximation (CMF-10), we have 
three choices of clusters shown in Fig.~\ref{figS1}. 
While we treat exactly the interactions within the cluster, 
the interactions between the cluster and the rest of the system 
are also included via effective fields acting at the cluster edge.
The effective fields are determined self-consistently via 
the expectation values of the operators 
$\langle \hat{a}_j\rangle$ and $\langle \hat{n}_j\rangle$, 
in a standard manner~\cite{CMFa}.

In the main text, we have performed a cluster-size scaling of the CMF results for the amplitude of a discontinuous jump $\Delta_1$ at the transition between the two SS phases [see, Fig. 1(d)]. 
To extrapolate the value of $\Delta_1$, we introduced the scaling parameter $\lambda$ defined by $N_{\rm B}/(N_{\rm C}\times z/2)$, which varies from $0$ to $1$. Here, $N_{\rm B}$ is the number of bonds within the cluster and $z$ is the coordination number for the lattice, i.e., $z=6$ in this case. The denominator means the number of bonds of the original lattice per $N_{\rm C}$ sites, and hence the parameter $\lambda$ provides an indication of how much the correlation effects between particles are taken into account in the cluster. The value of $\lambda$ for each cluster is shown in Table~\ref{table1}. 
In Fig. 1(d) we performed a linear extrapolation toward $\lambda=1$ using the three points of CMF-6, -10, and -15, since the three-site cluster is too small for scaling. The linear fits of these three points are fairly good, as seen in Fig. 1(d). At $J/V=0.108$, the jump $\Delta_1$ vanishes in the limit $\lambda \rightarrow 1$, which means that the triple point of the two SS phases and the SF phase is strongly reduced from the MF value $J/V=0.25$~\cite{murthy-97a} to $J/V=0.108$ due to the quantum fluctuations. This estimate for the triple point is in fairly good agreement with the recent QMC prediction~\cite{bonnes-11a}.

\subsection{\label{2}B. Additional data for the transition between two SS phases} 
In the main text, we made only a passing reference to the transition between 
the high-density SS phase (often called SSA for distinction) and the low-density SS phase (SSB), since the main focus was the hysteresis occurring near the transition between solid (or SS) and SF phases. 
We present here some additional results for the properties of the SSA-SSB transition.

\begin{figure}[t]
\includegraphics[scale=0.26]{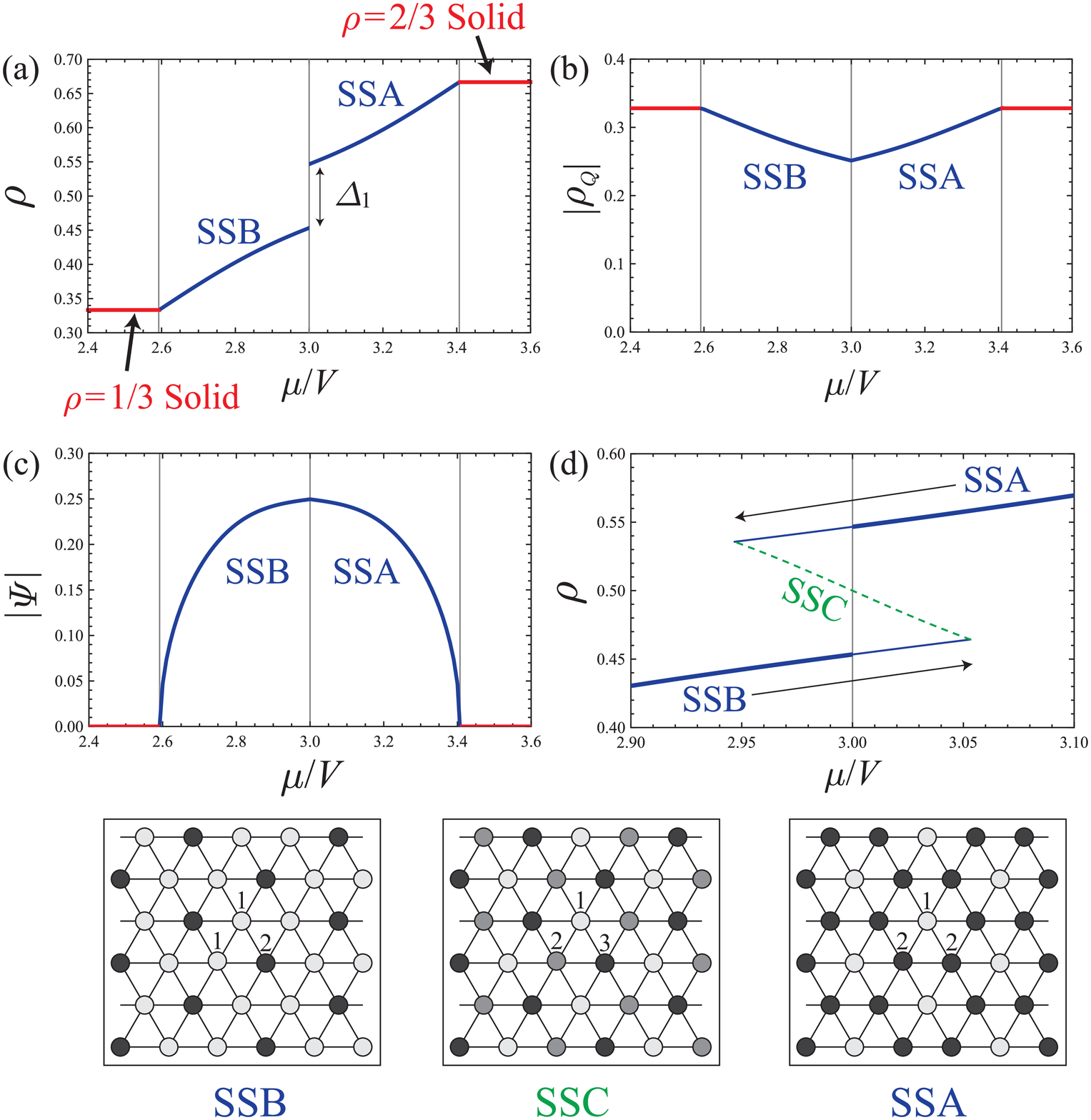}
\caption{\label{figS2}
The $\mu/V$ dependences of (a) the average density $\rho$, (b) the solid order parameter $|\rho_{\bf Q}|$, and (c) the SF order parameter $|\Psi|$ for $J/V=0.1$, obtained by the CMF-10 calculations. 
(d) Magnified view of the region near $\mu/V=3$ in (a). The thick solid, thin solid, and dashed lines represent the solutions of the ground, metastable, and unstable states, respectively. The lower panels show the sublattice structures of SSA, SSB, and SSC. 
}
\end{figure}

In Figs.~\ref{figS2}(a-d), the results of the CMF-10 for the filling factor $\rho$, the solid order parameter $|\rho_{\bf Q}|$ with ${\bf Q}=(4\pi/3d,0)$, and the SF order parameter $|\Psi|$ are shown as functions of $\mu/V$ for $J/V=0.1$. 
As shown in Fig.~\ref{figS2}(a), the ground-state density curves of SSA and SSB are not connected at the center, while the two order parameters in Figs.~\ref{figS2}(b) and (c) do not show such discontinuity. 
The latter is because of the particle-hole symmetry at $\mu/V=3$. 
The discontinuity in the density indicates the first-order character of the transition between the two SS phases, although the amplitude of the jump is strongly reduced in the limit $\lambda \rightarrow 1$ [see, Fig. 1(d)].

In addition, we plot in Fig.~\ref{figS2}(d) the solution curve of the CMF-10 equation, including the metastable and unstable solutions, near the SSA-SSB phase boundary. 
The solution lines of SSA and SSB are extended due to the metastability, and connected by the line of the three-sublattice SS state, called SSC~\cite{wessel-05a,boninsegni-05a,sen-08a,heidarian-10a}. We found that unstable SSC solutions (corresponding to unstable stationary points of the free energy) exist in the vicinity of half filling, although there is no region of the SSC phase in the ground-state phase diagram. The density versus chemical potential curve has negative curvature (negative compressibility) in the SSC state, which indicates the emergence of phase separation at a fixed density near $\rho=1/2$; the system separates into a mixture of SSA and SSB phases.

Note that in the case of infinite-range dipole-dipole interactions, there is a finite region of stable SSC phase in the phase diagram (marked SS2 in Fig.~4), although it is remarkably narrow for CMF-10. It is an interesting future issue to determine whether the SSC phase is actually stable or not for the infinite-range interaction model.


\begin{thebibliography}{99}
\bibitem{griesmaier-05}
A. Griesmaier $et$ $al$., Phys. Rev. Lett. {\bf 94}, 160401 (2005).

\bibitem{lahaye-07}
T. Lahaye $et$ $al$., Nature (London) {\bf 448}, 672 (2007).

\bibitem{ni-08}
K.-K. Ni $et$ $al$., Science {\bf 322}, 231 (2008).

\bibitem{ospelkaus-09}
S. Ospelkaus $et$ $al$., Faraday Discuss. {\bf 142}, 351 (2009).

\bibitem{aikawa-10}
K. Aikawa $et$ $al$., Phys. Rev. Lett. {\bf 105}, 203001 (2010).

\bibitem{becker-10}
C. Becker $et$ $al$., New J. Phys. {\bf 12}, 065025 (2010). 

\bibitem{gardner-99}
J. S. Gardner $et$ $al$., Phys. Rev. Lett. {\bf 83}, 211 (1999).

\bibitem{balents-10}
L. Balents, Nature (London) {\bf 464}, 199 (2010). 

\bibitem{bramwell-01}
S. T. Bramwell $et$ $al$., Phys. Rev. Lett. {\bf 87}, 047205 (2001).

\bibitem{matsuda-70}
H. Matsuda and T. Tsuneto, Suppl. Prog. Theor. Phys. 
{\bf 46}, 411 (1970). 

\bibitem{pollet-10}
L. Pollet $et$ $al$., Phys. Rev. Lett. {\bf 104}, 125302 (2010).

\bibitem{oguchi-55}
T. Oguchi, Prog. Theor. Phys. {\bf 13}, 148 (1955).

\bibitem{hassan-07}
S. R. Hassan, L. de Medici, and A.-M. S. Tremblay, Phys. Rev. B {\bf 76}, 144420 (2007). 

\bibitem{yamamoto-09} 
H. A. Bethe, Proc. R. Soc. London, Ser. A {\bf 150}, 552 (1935); D. Yamamoto, Phys. Rev. B {\bf 79}, 144427 (2009). 

\bibitem{EPAPS} 
See the supplementary material attached below.

\bibitem{murthy-97}
G. Murthy, D. Arovas, and A. Auerbach, Phys. Rev. B {\bf 55}, 3104 (1997).

\bibitem{boninsegni-05heidarian-05sen-08heidarian-10}
M. Boninsegni and N. Prokof'ev, Phys. Rev. Lett. {\bf 95}, 237204 (2005); 
D. Heidarian and K. Damle, Phys. Rev. Lett. {\bf 95}, 127206 (2005);
R. G. Melko $et$ $al$., Phys. Rev. Lett. {\bf 95}, 127207 (2005);
A. Sen $et$ $al$., Phys. Rev. Lett. {\bf 100}, 147204 (2008);
D. Heidarian and A. Paramekanti, Phys. Rev. Lett. {\bf 104}, 015301 
(2010). 

\bibitem{wessel-05}
S. Wessel and M. Troyer, Phys. Rev. Lett. {\bf 95}, 127205 (2005).

\bibitem{Note} After the release of the preprint version of this Letter, this prediction has been confirmed by QMC [L. Bonnes and S. Wessel, Phys. Rev. B {\bf 84}, 054510 (2011); X.-F. Zhang $et$ $al$., $ibid$. {\bf 84}, 174515 (2011)]. 

\bibitem{sansone-10danshita-10}
B. Capogrosso-Sansone $et$ $al$., Phys. Rev. Lett. {\bf 104}, 125301 
(2010); 
I. Danshita and D. Yamamoto, Phys. Rev. A {\bf 82}, 013645 (2010). 

\bibitem{batrouni-00}
G. G. Batrouni and R. T. Scalettar, Phys. Rev. Lett. {\bf 84}, 1599 
(2000). 

\bibitem{mag}
K.-K. Ng and T. K. Lee, Phys. Rev. Lett. {\bf 97}, 127204 (2006); P. Sengupta and C. D. Batista, Phys. Rev. Lett. {\bf 98}, 227201 (2007). 

\end{thebibliography}

\begin{thebibliography}{99}
\bibitem{batrouni-00a}
G. G. Batrouni and R. T. Scalettar, Phys. Rev. Lett. {\bf 84}, 1599 
(2000). 

\bibitem{CMFa}
T. Oguchi, Prog. Theor. Phys. {\bf 13}, 148 (1955);
H. H. Chen and F. Lee, Phys. Rev. B {\bf 48}, 9456 (1993);
A. J. Garc\'ia-Adeva and D. L. Huber, Phys. Rev. Lett. {\bf 85}, 4598 (2000);
S. R. Hassan, L. de Medici, and A.-M. S. Tremblay, Phys. Rev. B {\bf 76}, 144420 (2007).

\bibitem{murthy-97a}
G. Murthy, D. Arovas, and A. Auerbach, Phys. Rev. B {\bf 55}, 3104 (1997).

\bibitem{bonnes-11a}
L. Bonnes and S. Wessel, Phys. Rev. B {\bf 84}, 054510 (2011). 

\bibitem{boninsegni-05a}
M. Boninsegni and N. Prokof'ev, Phys. Rev. Lett. {\bf 95}, 237204 (2005). 

\bibitem{wessel-05a}
S. Wessel and M. Troyer, Phys. Rev. Lett. {\bf 95}, 127205 (2005).

\bibitem{sen-08a}
A. Sen, P. Dutt, K. Damle and R. Moessner, Phys. Rev. Lett. {\bf 100}, 147204 (2008). 

\bibitem{heidarian-10a}
D. Heidarian and A. Paramekanti, Phys. Rev. Lett. {\bf 104}, 015301 (2010).

\end{thebibliography}
\end{document}